\documentclass[a4paper,10pt,conference]{IEEEtran}

\usepackage[USenglish]{babel}
\usepackage{float}
\usepackage{graphicx}
   \DeclareGraphicsExtensions{.eps}
\usepackage{subfig}
\usepackage{amsmath}
\usepackage{amsthm}
\usepackage{amsfonts}
\usepackage{multirow}
\usepackage{algorithm}
\usepackage{algorithmic}

\newcommand{\code}{\mathcal{C}}

\newcommand{\pth}{p_{\text{th}}}
\newcommand{\f}{\mathbb{F}}

\newcommand{\inef}{\mu}
\newcommand{\inefm}{\bar{\mu}}
\newcommand{\ainef}{\mu_{\text{th}}}

\newcommand{\inefpeg}{\inefm_{\text{PEG}}}
\newcommand{\inefmpeg}{\inefm_{\text{ModPEG}}}
\newcommand{\inefspeg}{\inefm_{\text{SPEG}}}
\newcommand{\inefrand}{\inefm_{\text{RAND}}}

\newcommand{\stdpeg}{\sigma_\text{{PEG}}}
\newcommand{\stdmpeg}{\sigma_\text{{ModPEG}}}
\newcommand{\stdspeg}{\sigma_\text{{SPEG}}}

\newcommand{\ee}{E}
\newcommand{\vs}{S}

\newcommand{\eij}{(c_i,s_j)}
\newcommand{\hij}{h_{i,j}}

\newcommand{\dm}{d_{\text{max}}}
\newcommand{\vsd}{S_{d}}

\newcommand{\ndi}{n_d^{(i)}}

\newcommand{\ndt}{n_d^{(t)}}

\newcommand{\vsdt}{S_{d}^{(t)}}

\newcommand{\fsdi}{f_{d}^{(i)}}
\newcommand{\fsdt}{f_{d}^{(t)}}

\newcommand{\inputdata}{\left(n,m,D_s\right)}
\newcommand{\inputdataSPEG}{\left(n,m,\left\{\vsdt\right\}\right)}
\newcommand{\inputdataSPEGopt}{\left(n,m,\left\{\fsdt\right\}\right)}

\newcommand{\peg}{\mathcal{G}_{\text{PEG}}\inputdata}
\newcommand{\mpeg}{\mathcal{G}_{\text{MPEG}}\inputdata}
\newcommand{\speg}{\mathcal{G}_{\text{SPEG}}\inputdataSPEG}
\newcommand{\spegopt}{\mathcal{G}_{\text{SPEG}}\inputdataSPEGopt}

\theoremstyle{definition}

\theoremstyle{definition}

\newtheorem{expl}{Example}[section]
\theoremstyle{remark}

\IEEEoverridecommandlockouts
\title{Scheduled-PEG construction of LDPC codes for Upper-Layer FEC}
\author{%
  Lam Pham Sy$^{\flat}$, Valentin Savin$^{\natural}$, David Declercq$^{\sharp}$, Nghia Pham$^{\flat}$ \\%
  {\small $^{\flat}$Eutelsat, Paris, France,  $^{\natural}$CEA-LETI, MINATEC campus, Grenoble, France,  $^{\sharp}$ENSEA, Cergy-Pontoise, France}%
  \thanks{This work was supported by the French National Research Agency (ANR), grant No 2009 VERS 019 04 -- ARSSO project.}
}

\begin{document}
\maketitle
\begin{abstract}
The Progressive Edge Growth (PEG) algorithm is one of the most widely-used method for constructing finite length LDPC
codes. In this paper we consider the PEG algorithm together with a {\em scheduling distribution}, which specifies the
order in which edges are established in the graph. The goal is to find a scheduling distribution that yields ``the
best'' performance in terms of {\em decoding overhead}, performance metric specific to erasure codes and widely used
for upper-layer forward error correction (UL-FEC). We rigorously formulate this optimization problem, and we show that
it can be addressed by using genetic optimization algorithms. We also exhibit PEG codes with optimized scheduling
distribution, whose decoding overhead is less than half of the decoding overhead of their classical-PEG counterparts.
\end{abstract}
  \section{Introduction}\label{sec:introduction}
Data loss recovery -- for instance, for content distribution applications or for distributed storage systems -- is
widely addressed using erasure codes that operate at the transport or the application layer of the communication
system. These codes, referred to as upper-layer (UL) codes, extend source data packets with repair (redundant) packets,
which are used to recover the lost data at the receiver. They are generally proposed in conjunction with physical layer
codes, in order to maximize the reliability of the transmission system, especially in case of intermittent connectivity
or deep fading of the signal for short periods.
%
%
In such situations, the physical layer FEC fails and we can either ask for retransmission (only if a return channel
exists, and penalizing in broadcast/multicast scenarios) or use UL-FEC. Hence, the use of UL-FEC codes is of critical
importance in broadcast communication systems in general, and satellite communications in particular.
%

Low Density Parity Check (LDPC) codes constitute a very broad class of FEC codes, distinguished by the fact that they
are defined by sparse parity-check matrices, and can be iteratively decoded in linear time with respect to their
block-length. Invented by Gallager in early 60's \cite{gall-monograph}, but considered impractical to implement, these
codes have been neglected for more that three decades, and ``rediscovered'' in the late 90's \cite{MacK}. Nowadays, a
large body of knowledge has been acquired (analysis, optimization, construction); LDPC codes are known to be capacity
approaching codes for a large class of channels \cite{Rich-Shok-Urba}, and became synonymous with modern coding.

However, this capacity approaching property holds in the asymptotic limit of the code length, and codes optimized from
this asymptotic perspective may suffer significant performance degradation at practical lengths. Actually, the
asymptotic optimization, performed by using density-evolution methods \cite{Rich-Urba}, yields an {\em irregularity
profile}, which specifies the distribution of node-degrees in the bipartite (Tanner) graph \cite{Tann} associated with
the code. It is assumed that the girth\footnote{Length of a shortest cycle.} of the bipartite graph goes to infinity
with the code-length. Hence, optimized irregularity profiles can be used to construct codes that are ``long enough''
(at least few thousand bits) to avoid short cycles, although they must be ``short enough'' to be practical.

One of the most widely-used method for constructing finite length codes is the Progressive Edge Growth (PEG) algorithm
\cite{PEG}. It constructs bipartite graphs with large girth, by establishing edges progressively: the graph grows in an
edge-by-edge manner, optimizing each local girth. There is an {\em underlying edge order} within the PEG, corresponding
to the order in which edges are established in the graph. In general, edges are progressively established starting with
those incident to symbol-nodes of degree-$2$ and ending with those incident to symbol-nodes of maximum degree. However,
any other order with respect to the symbol-node degrees would also be possible. Besides, for a given symbol-node
degree, edges can be established in a {\em node-by-node} manner (all edges incident to some symbol node are established
before moving to the next symbol-node), or in a {\em degree-by-degree} manner (a first edge is established for each
symbol-node, then a second edge is established for each symbol-node, and so on until all the symbol-nodes reach the
given degree). Although this order may significantly impact the performance of the constructed code, it is rather
difficult to formalize and has practically not been investigated in the literature. There are however several papers
that aim to enhance the PEG construction by optimizing some objective function, as for instance minimizing the number
of cycles created \cite{Venkiah08}, or minimizing the approximate cycle extrinsic (ACE) message degree
\cite{Tian04:ACE}, \cite{Xiao04}.

In this paper we consider the PEG algorithm together with a {\em scheduling distribution}, which will be referred to as
scheduled-PEG, or SPEG for short. Within the SPEG algorithm, symbol-nodes are divided into subsets, each subset
containing symbol-nodes of same degree. Edges incident to the symbol-nodes of a subset are established in a
degree-by-degree manner, before moving to the next subset. The scheduling distribution specifies the fraction of nodes
within each subset. Our purpose is to find a scheduling distribution that yields the best performance in terms of
decoding overhead (performance metric widely used for UL-FEC). We rigorously formulate this optimization problem, and
we show that it can be addressed by using genetic optimization algorithms.

The paper is organized as follows. Section \ref{sec:Defi_Nots} gives a brief overview of the basic theory and
definitions related to LDPC codes, their iterative decoding, and the associated performance metrics over the BEC. The
construction of finite length LDPC codes is addressed in Section \ref{sec:Constr_FL}. The proposed Scheduled-PEG
algorithm is also introduced in this section. Section \ref{sec:Opt_SPEG} focuses on the optimization of the Scheduled
PEG algorithm and presents simulation results. Finally, Section \ref{sec:Conclusion} concludes the paper.

 \section{LDPC Codes and Performance Metric over the Erasure Channel}\label{sec:ldpc_codes}\label{sec:Defi_Nots}

\subsection{Binary and non-binary LDPC codes}

In this paper we consider both binary and non-binary LDPC codes defined over some finite field $\f_q$, with $q=2^p$
\cite{Davey-MacKey}. If $p=1$, the code is binary. We fix, once for all, a vector space isomorphism:
\begin{equation}
\f_q \tilde{\longrightarrow} \f_2^p \label{eq:vector_isomors_def}
\end{equation}
Elements of $\f_q$ will be called symbols. We say that the binary sequence $(x_0,..,x_{p-1}) \in \f_2^p$ is the
\textit{binary image} of the symbol $X \in \f_q$, iff they correspond to each other by the above isomorphism.

An LDPC code is a linear code defined by a sparse parity-check matrix $H\in{\mathbb{M}}_{m,n}(\f_q)$. Alternatively, it
can be represented by a bipartite (Tanner) graph\footnote{By abusing language, throughout the paper, the term ``code''
will be used with the meaning of ``bipartite graph''.} \cite{Tann}, containing $n$ symbol-nodes and $m$
constraint-nodes associated respectively with the $n$ columns and $m$ rows of  $H$. A symbol-node and a constraint-node
are connected by an edge if and only if the corresponding entry of $H$ is non-zero, in which case the edge is assumed
to be ``labeled'' by the non-zero entry.

Symbol-nodes take values in $\f_q$, and a constraint-node is said to be verified if the linear combination of neighbor
symbols (with coefficients given by the corresponding edge labels) is equal to zero. A non-binary word
$(X_1,\dots,X_n)$ is a codeword if it verifies all the constraint nodes of the graph.

The degree of a node is by definition the number of edges incident to that node (number of non-zero entries on the
corresponding row/column of $H$). A code is called {\em $(d_s, d_c)$-regular} if all symbol-nodes are of degree $d_s$
and all constraint-nodes are of degree $d_c$; otherwise it is called {\em irregular}.

Let $\Lambda_d$ and $\Gamma_d$ denote respectively the fractions of symbol and constraint nodes of degree-$d$. Let also
$\lambda_d$ and $\rho_d$ be the fractions of edges connected respectively to symbol and constraint nodes of degree-$d$.
The degree distribution polynomials, from the node and the edge perspective, are defined by:
$$\begin{array}{r@{\ }lr@{\ }ll}
  \Lambda(x) = & \displaystyle\sum_d \Lambda_d x^{d},   & \ \Gamma(x) = & \displaystyle\sum_d \Gamma_d x^{d} & \mbox{\footnotesize(node-persp.)} \\
  \lambda(x) = & \displaystyle\sum_d \lambda_d x^{d-1}, & \   \rho(x) = & \displaystyle\sum_d \rho_d x^{d-1} & \mbox{\footnotesize(edge-persp.)}
\end{array}$$

The designed code rate, denoted by $r$, is by definition:
 $$r = 1 - \displaystyle\frac{\int_{0}^{1}\rho(x)\text{d}x}{\int_{0}^{1}\lambda(x)\text{d}x} = 1 - \displaystyle\frac{\Lambda '(1)}{\Gamma'(1)}$$

If the parity-check matrix is of rank $m$, then the (non-binary) code dimension is equal to $k = n-m$, and $r$ is equal
to the code rate, that is $r = \frac{k}{n}$.

We also denote by $K$ and $N$ the {\em binary code dimension} and the {\em binary code length}, respectively. Hence, $K
= kp$ and $N = np$.

\subsection{Iterative erasure decoding}

For binary LDPC codes over the BEC, the belief-propagation (BP) decoding translates into a simple technique of
recovering the erased bits, by iteratively searching for check-nodes with only one erased neighbor bit-node
\cite{zyablov}. Indeed, if a check-node is connected to only one erased bit-node, the value of the latter can be
recovered as the XOR of all the other neighbor bit-nodes. This value is then injected into the decoder, and the
decoding continues by searching for some other check-node having a single erased neighbor bit-node. We remark that the
erasure decoding can be performed on-the-fly: decoding starts as soon as the first bit is received and each new
received bit is injected on-the-fly into the decoder. The decoding will stop by itself if all the bits have been
recovered, or when it ``gets stuck'' because any check-node is connected to at least two erased bit-nodes (such a
configuration is called a {\em stopping set}).

The above considerations can also be generalized in the case of non-binary codes. It is important to note that we
consider {\em non-binary LDPC codes over a binary erasure channel}, which means that the channel erases bits from the
binary image of the transmitted codeword. Hence, a coded symbol can be completely erased (all the bits of its binary
image are erased), completely received (no bit of its binary image is erased), or partially erased/received (some bits
of its binary image are erased, some others are received). At the receiver part, the received bits are used to
(partially) reconstruct the corresponding symbols of the transmitted codeword. Thus, for each symbol node we can
determine the set of {\em eligible symbols}, i.e. symbols whose binary images match the received bits. Such a set
contains only one symbol (namely the transmitted one) if the corresponding symbol-node is completely received. These
sets are then iteratively updated, according to the linear constraints between symbol-nodes \cite{savin2008nbl}.
Alternatively, non-binary LDPC codes can be decoded by using their extended binary image~\cite{savin:blte}.

\subsection{Performance metrics for finite-length codes}\label{subsec:perf_metrics}
A performance metric that is often associated with on-the-fly decoding is the {\em decoding inefficiency}, defined as
the ratio between the number of received bits when decoding completes and the number of information bits
\cite{Byers98}. More precisely, we assume that the encoded bit-stream is permuted according to some random permutation
$\pi$. The permuted bit-stream is sequentially delivered to the decoder, which performs erasure decoding
on-the-fly\footnote{Such a random reception corresponds to a randomly interleaved erasure channel, which allows us to
dispense with a specific loss model.}: each bit is injected into the decoder in the appropriate position and erasure
decoding is performed until either decoding completes or it gets stuck. If decoding gets stuck, we inject the next bit
from the permuted bit-stream. We denote by $k_\pi$ the number of bits from the permuted bit-stream that have been
injected into the decoder when the erasure decoding completes; the value $k_\pi-k$ is referred to as {\em reception
overhead}. The decoding inefficiency is defined as $\inef(\pi) = \frac{k_\pi}{k}$ . It is a random variable, whose
value depends on the (random) permutation $\pi$. Note also that $\inef(\pi)\in\left[1, \frac{1}{r}\right]$, where $r$
is the code rate. The average decoding inefficiency is defined as the expected value of $\inef$; that is:
  $$ \inefm = {\text{E}}\left[\inef\right] = \frac{1}{n!}\sum_{\pi}\inef(\pi)$$
Note that $\inefm = 1$ if and only if the code is MDS (its minimum distance is equal to $n-k+1$). More accurate
statistics about the decoding inefficiency are provided by the {\em probability of decoding failure}, defined as the
complementary cumulative distribution function (CCDF) of $\inef$:
  $$F(x) = \Pr[ \inef(\pi) > x ], \ \ x\in\left[1, \frac{1}{r}\right]$$
Hence, $F(x)$ is the probability of decoding failure assuming that the number of bits received from the channel is
equal to $kx$, or equivalently, the reception overhead is equal to $k(x-1)$. Indeed, the $kx$ bits received from the
channel are the first $kx$ bits of the encoded bit-stream permuted by some permutation $\pi$. Hence, a decoding failure
occurs if and only if $\inef(\pi) > x$.

From the above definition, it also follows that:
 $$\int_{1}^{\frac{1}{r}} F(x) \ \mbox{d}x = \inefm - 1$$
Therefore, if $\code_1$ and $\code_2$ are two codes such that $\inefm(\code_1) < \inefm(\code_2)$, then $\code_1$
presents a better performance in the waterfall region of $F$, but this might happen at the expense of a higher error
floor\footnote{The waterfall is the region in which the failure probability decreases very quickly as $x$ increases.
However, there might be a point after which the curve does not fall as quickly as before, in other words, there is a
region in which performance flattens. This region is called the error floor region.}.

Finally, we note that the probability of decoding failure can also be expressed in relation with the fraction of erased
bits (rather than the fraction of received bits); in this case it will be referred to as {\em Frame Error Rate} (FER).
The {\em Bit Error Rate} (BER) will denote the probability of a bit being erased (after the decoding process), assuming
that a certain fraction of bits have been received.

\subsection{Asymptotic performance}
Unlike the finite-length performance, the asymptotic performance does not refer to the performance of a given code, but
to the performance of a given family or {\em ensemble of codes}. Such an ensemble contains arbitrary-length codes that
share the same properties in terms of distributions of node-degrees in the associated bipartite graph.

Let $E(\lambda,\rho)$  denote the ensemble of LDPC codes of arbitrary length $n>0$, with edge perspective
degree-distributions polynomials  $\lambda$ and  $\rho$.  When $n$ goes to infinity, (almost) all the codes behave
alike, and they exhibit a threshold phenomenon, separating the region where reliable transmission is possible from that
where it is not \cite{Rich-Urba}. Assume that an arbitrary code $\code_n \in E(\lambda,\rho)$, of length  $n$, is used
over the BEC, and let $p_e$ denote the channel erasure probability. The threshold of the ensemble $E(\lambda,\rho)$ is
defined as the supremum value of $p_e$ (i.e. the worst channel condition) that allows transmission with an arbitrary
small error probability, assuming that $n$ goes to infinity. Let us denote this threshold value by
$p_{\mbox{\scriptsize th}}(\lambda,\rho)$. The threshold value is necessarily less than the channel capacity, that is
$p_{\mbox{\scriptsize th}}(\lambda,\rho) \leq 1-r$, where $r$ is the (asymptotic) code rate of the ensemble
$E(\lambda,\rho)$. Roughly speaking, this means that if an encoded sequence of length $n$ is transmitted over the
channel, it can be successfully decoded iff the fraction of erased bits is less than  $p_e < p_{\mbox{\scriptsize
th}}(\lambda,\rho)$, with $p_e \rightarrow  p_{\mbox{\scriptsize th}}(\lambda,\rho)$ as  $n\rightarrow +\infty$. It is
assumed here that the girth of the graph goes to infinity with $n$, which actually happens for almost all the codes in
$E(\lambda,\rho)$. It follows that the decoding inefficiency, which can be expressed as $\frac{(1-p_e)N}{K} =
\frac{1-p_e}{r}$, also goes to a threshold value:
   $$\ainef = \frac{1-\pth}{r},$$
which will be referred to as {\em inefficiency threshold}.

Given an ensemble  $E(\lambda,\rho)$, its threshold value can be efficiently computed by tracking the fraction of
erased messages passed during the belief propagation decoding. This method is called density evolution (the name is due
to the fact that over more general channels, we have to track the message densities). For more details on density
evolution we refer to \cite{Rich-Urba} for binary codes, and \cite{rathi:det} and \cite{savin2008nbl} for non-binary
LDPC codes. The introduction of irregular codes, as well as the asymptotic optimization based on the density evolution
method, made possible the construction of capacity approaching ensembles of LDPC codes \cite{Rich-Shok-Urba}.

 \section{Finite length LDPC codes construction}\label{sec:Constr_FL}
As discussed in the above section, the asymptotic threshold can be approached by long codes, which do not contain short
cycles. Short cycles may also harm the performance of short (finite-length) codes, as they can result in short stopping
sets. Hence, the PEG algorithm has been proposed, and is widely used, for constructing bipartite graphs with large
girth, in a best effort sense, by progressively establishing edges between symbol and check nodes in an edge-by-edge
manner~\cite{PEG}.

\subsection{Progressive Edge Growth algorithm}

A bipartite (Tanner) graph is denoted as $(S, C, E)$, where $S = \left\{s_1,s_2,...,s_{n}\right\}$ is the set of
symbol-nodes, $C=\left\{c_1,c_2,...,c_{m}\right\}$ is the set of constraint nodes and $E \subseteq S \times C$ is the
set of edges. An edge $\eij \in E$ corresponds to a non-zero entry $\hij$ of the parity check matrix $H$. We also
denote by $D_S$ the ``target sequence'' of symbol-node degrees, which is assumed to be sorted in non-decreasing order:
  $$ D_S = \left\{ d_{s_1}, d_{s_2},..., d_{s_{n}} \left| d_{s_1} \leq d_{s_2} \leq ... \leq d_{s_{n}} \right.\right\}$$
where $d_{s_j}$ is the degree of symbol node $s_j$.

When the PEG algorithm starts, the set of edges is empty, $E = \emptyset$. Edges will be progressively added to $E$, as
explained shortly. Given a symbol node $s_j$, we denote by $\underline{C}_{s_j}$ the set of constraint-nodes whose
distance to $s_j$ is maximum, in the current settings; that is, given the current set of edges. The distance between
two nodes is the length of the shortest path connecting them. If there is no path between $s_j$ and some constraint
node $c_j$, the distance between them is set to $+\infty$. Hence, if $\ee_{s_j} = \emptyset$, the distance from $s_j$
to any constraint-node is $+\infty$ and $\underline{C}_{s_j} = C$, the set of all the constraint-nodes. If $\ee_{s_j}
\neq \emptyset$, $\underline{C}_{s_j}$ can be determined by expanding a subgraph from symbol node $s_j$ up to the
maximal depth (see \cite{PEG}).

Finally, we use $c_i \longleftarrow \left\{\underline{C}_{s_j} \mid \mbox{min deg}\right\}$ to denote a random
constraint node $c_i \in \underline{C}_{s_j}$, having the lowest degree (given the current set of edges of the graph).
The PEG algorithm can be summarized as follows:

\begin{algorithm}
\caption{Progressive Edge Growth algorithm}
\label{algo:PEG_ORG}
\begin{algorithmic}
\FOR{$j$ = 1 \TO $n$}
    \FOR{$k$ = 1 \TO $d_{s_j}$}
           \STATE Determine $\underline{C}_{s_j}$, given the current $E$;
           \STATE $c_i \longleftarrow \left\{\underline{C}_{s_j} \mid \mbox{min deg}\right\}$;
           \STATE Add edge $(s_j, c_i)$ to $E$;
    \ENDFOR
\ENDFOR
\end{algorithmic}
\end{algorithm}

We note that edges are established in a {\em node-by-node} manner, meaning that all edges incident to some symbol-node
are established before moving to the next symbol-node. We have further assumed that symbol nodes are sorted in
increasing order with respect to their degrees. Finally, we observe that the constraint-node degree distribution of the
constructed Tanner graph is almost uniform, {\em i.e.} all constraint nodes have only one or at most two consecutive
degrees \cite{PEG}.

Let $\peg$ denote the ensemble of all Tanner graphs constructed by using the PEG algorithm. The average inefficiency
ratio over all graphs in $\peg$ is defined as:
$$\inefpeg\inputdata = \text{E}\left[ \,\inefm(\code) \mid \code \in \peg \,\right]$$
When parameters $\inputdata$ are implied, it will be simply denoted by $\inefpeg$. The corresponding standard deviation
is denoted by $\stdpeg$.

\subsection{Modified Progressive Edge Growth algorithm}
The PEG graphs have large girth with respect to random graphs. Consequently, PEG graphs have low error floor in
comparison with random graphs. However, the error floor can be further lowered, by using a modification of the PEG
algorithm, called ModPEG.

Let $\vsd$ denote the set of symbol nodes of degree $d$, where $1 \leq d \leq \dm$ and $\dm$ denotes the maximum
symbol-node degree. Let $n_d$ denote number of symbol nodes within $\vsd$, hence $\sum_d n_d = n$.

Within the ModPEG algorithm, for each $\vsd$, edges are established in a {\em degree-by-degree} manner: a first edge is
established for each symbol-node, then a second edge is established for each symbol-node, and so on until all the
symbol-nodes in $\vsd$ reach the required degree $d$.

\begin{algorithm}
\caption{Modified Progressive Edge Growth algorithm}
\label{algo:PEG_VS}
\begin{algorithmic}
\FOR{$d$ = 1 \TO $\dm$}
    \FOR{$k$ = 1 \TO $d$}
        \FOR{$s_j \in \vsd$}
           \STATE Determine $\underline{C}_{s_j}$, given the current $E$;
           \STATE $c_i \longleftarrow \left\{\underline{C}_{s_j} \mid \mbox{min deg}\right\}$;
           \STATE Add edge $(s_j, c_i)$ to $E$;
        \ENDFOR
    \ENDFOR
\ENDFOR
\end{algorithmic}
\end{algorithm}

Let $\mpeg$ denote the ensemble of all Tanner graphs constructed by using the ModPEG algorithm. The average
inefficiency ratio over all graphs in $\mpeg$ is defined as:
$$\inefmpeg\inputdata = \text{E}\left[ \,\inefm(\code) \mid \code \in \mpeg \,\right]$$
When parameters $\inputdata$ are implied, it will be simply denoted by $\inefmpeg$. The corresponding standard
deviation is denoted by $\stdmpeg$.

\begin{table*}[!t]
\centering \caption{Optimized irregular LDPC codes, with rate $1/2$} \label{tab:opt_irr_distr}
\begin{tabular}{||c|l||r|r||r|r||r|r||}
\hline
Alphabet & Degree distributions & $\pth$ & $\ainef$ & $\inefpeg$ & $\stdpeg$ & $\inefmpeg$ & $\stdmpeg$\\
\hline
$\f_2$ & $\begin{array}{l} \Lambda(x) = 0.5489 x^2 + 0.2505 x^3 + 0.1608 x^7 +   0.0398 x^{30} \\ \Gamma(x) = 0.6609 x^8 +   0.3391 x^9 \end{array}$ & 0.4955 & 1.009 &  1.0829 & 1.475e-04 & 1.0876 & 1.202e-04\\
\hline
$\f_4$ & $\begin{array}{l} \Lambda(x) = 0.7140 x^2 + 0.2173 x^4 +   0.0687 x^{12} \\ \Gamma(x) = 0.7586 x^6 + 0.2414 x^7 \end{array}$ & 0.4926 & 1.0148 & 1.0604 & 2.903e-04 & 1.0685 & 2.117e-04 \\
\hline
$\f_8$ & $\begin{array}{l} \Lambda(x) = 0.7857 x^2 + 0.0529 x^3 +   0.1153 x^5 + 0.0461 x^{12} \\ \Gamma(x) = 0.2797 x ^5 + 0.7203 x ^6 \end{array}$ & 0.4931 & 1.0138 & 1.0655 & 3.188e-04 &   1.0726 & 2.8e-04 \\
\hline
$\f_{16}$ & $\begin{array}{l} \Lambda(x) = 0.8460   x^2 + 0.1056 x^5 + 0.0252   x^8 + 0.0232 x^{18} \\ \Gamma(x) = 0.3221 x^5 + 0.6779 x^6 \end{array}$ & 0.4945 & 1.011 &  1.0706 & 6.81e-04 & 1.0834 & 3.536e-04 \\
\hline
\end{tabular}
\end{table*}

\begin{figure*}[!t]
  \centering
  \subfloat[$\f_2$]{\label{fig:peg_avg_f2}\includegraphics[width=\columnwidth]{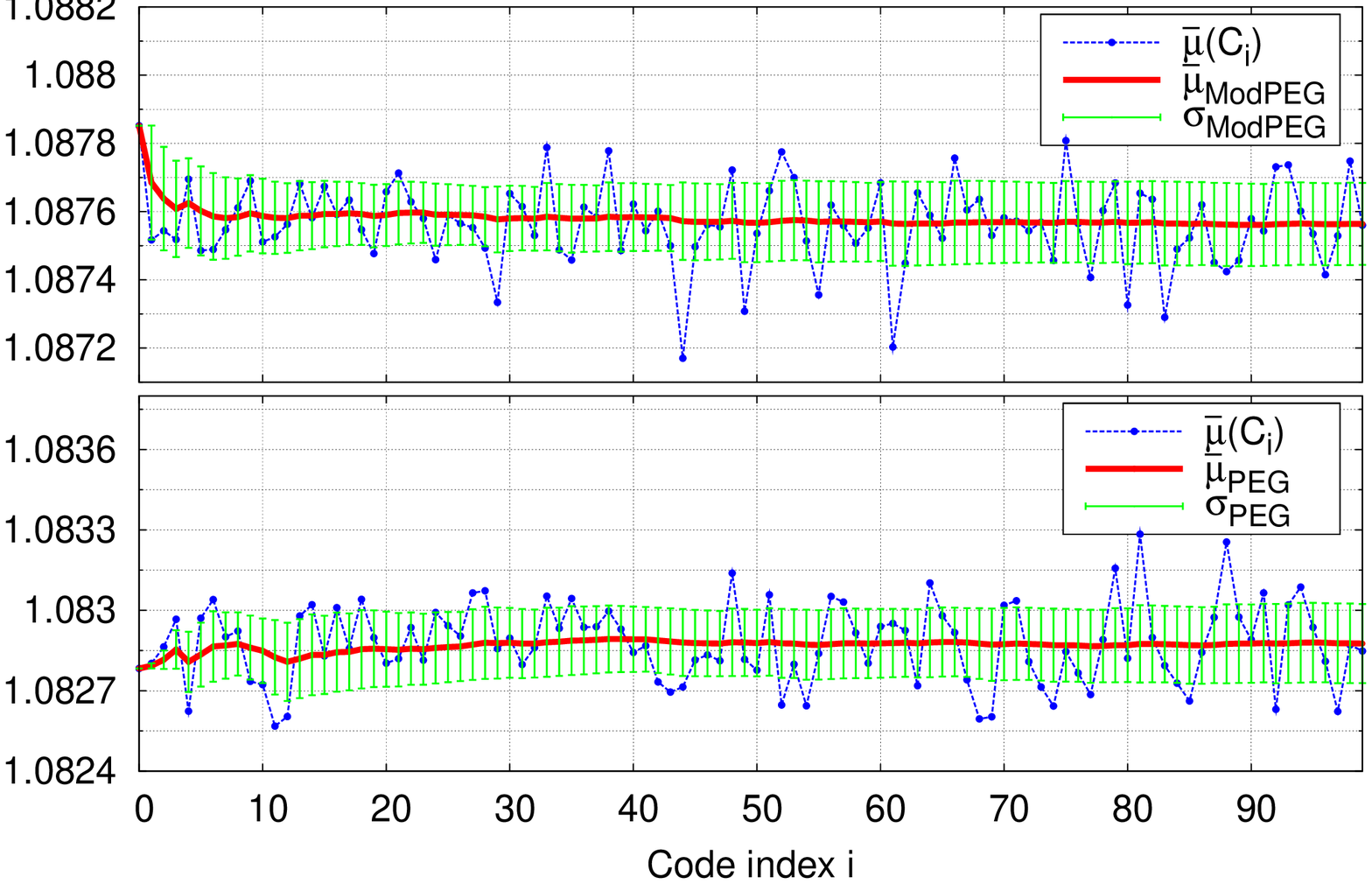}}
  \subfloat[$\f_4$]{\label{fig:peg_avg_f4}\includegraphics[width=\columnwidth]{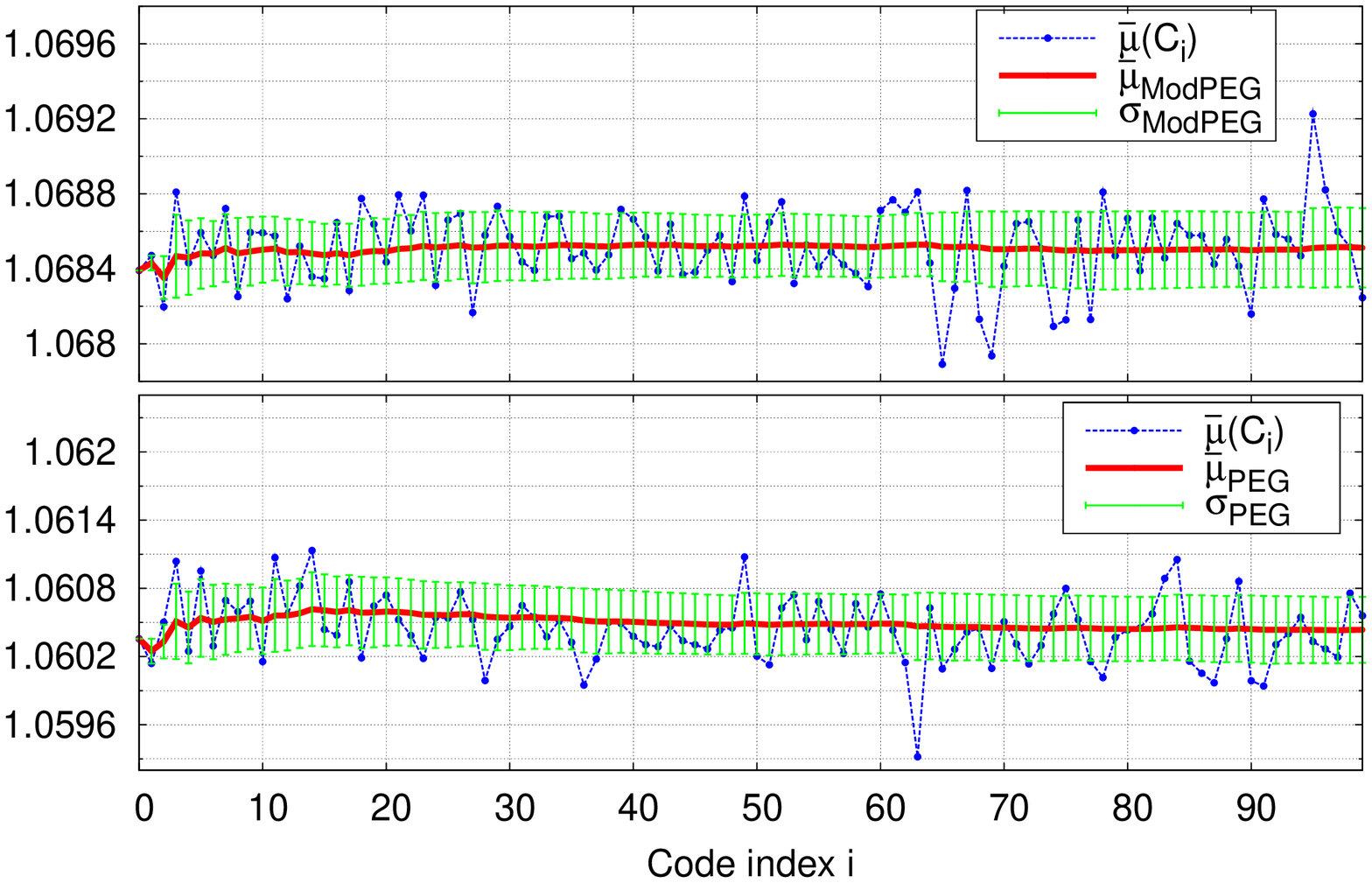}} \\
  \subfloat[$\f_8$]{\label{fig:peg_avg_f8}\includegraphics[width=\columnwidth]{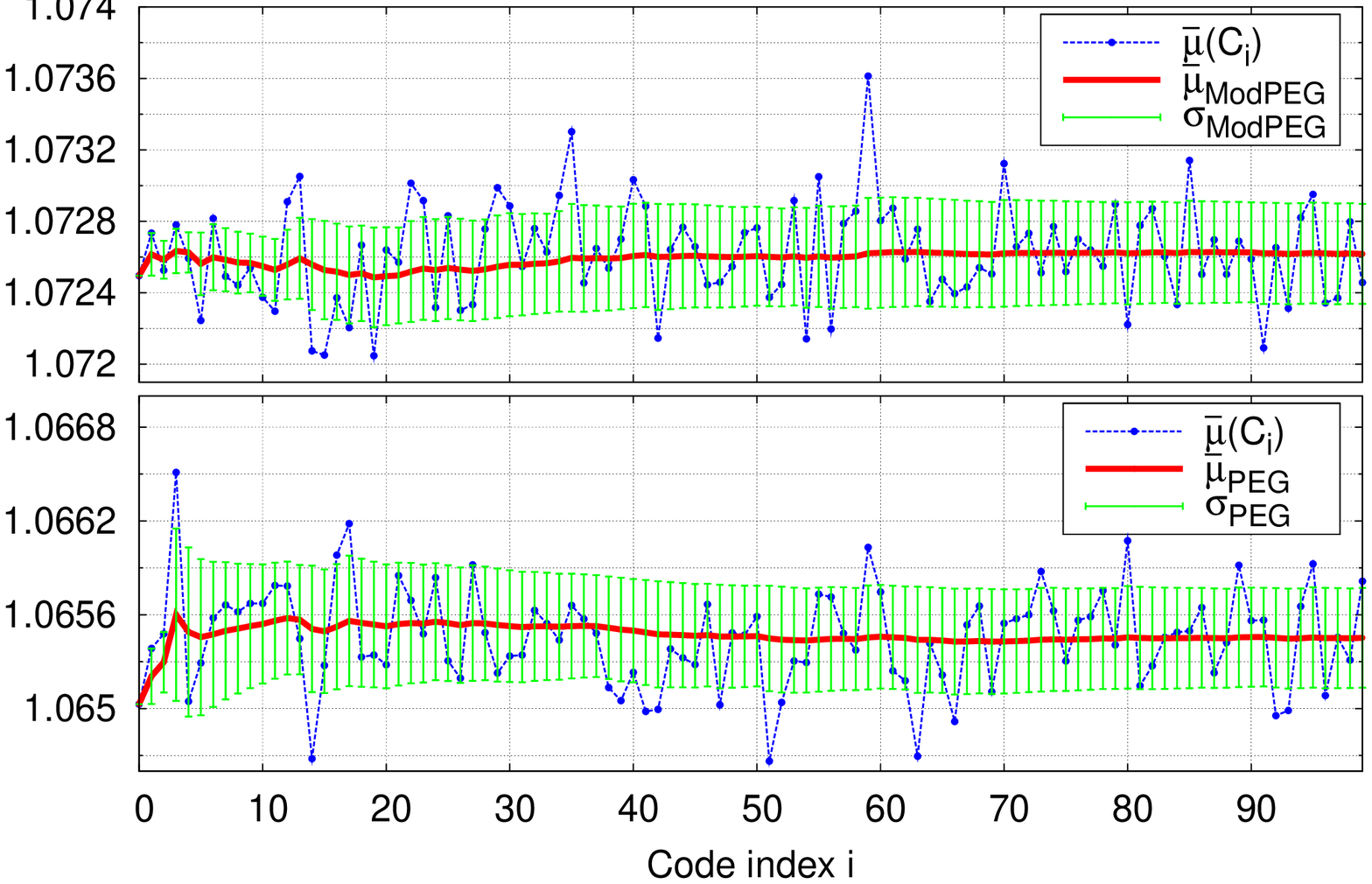}}
  \subfloat[$\f_{16}$]{\label{fig:peg_avg_f16}\includegraphics[width=\columnwidth]{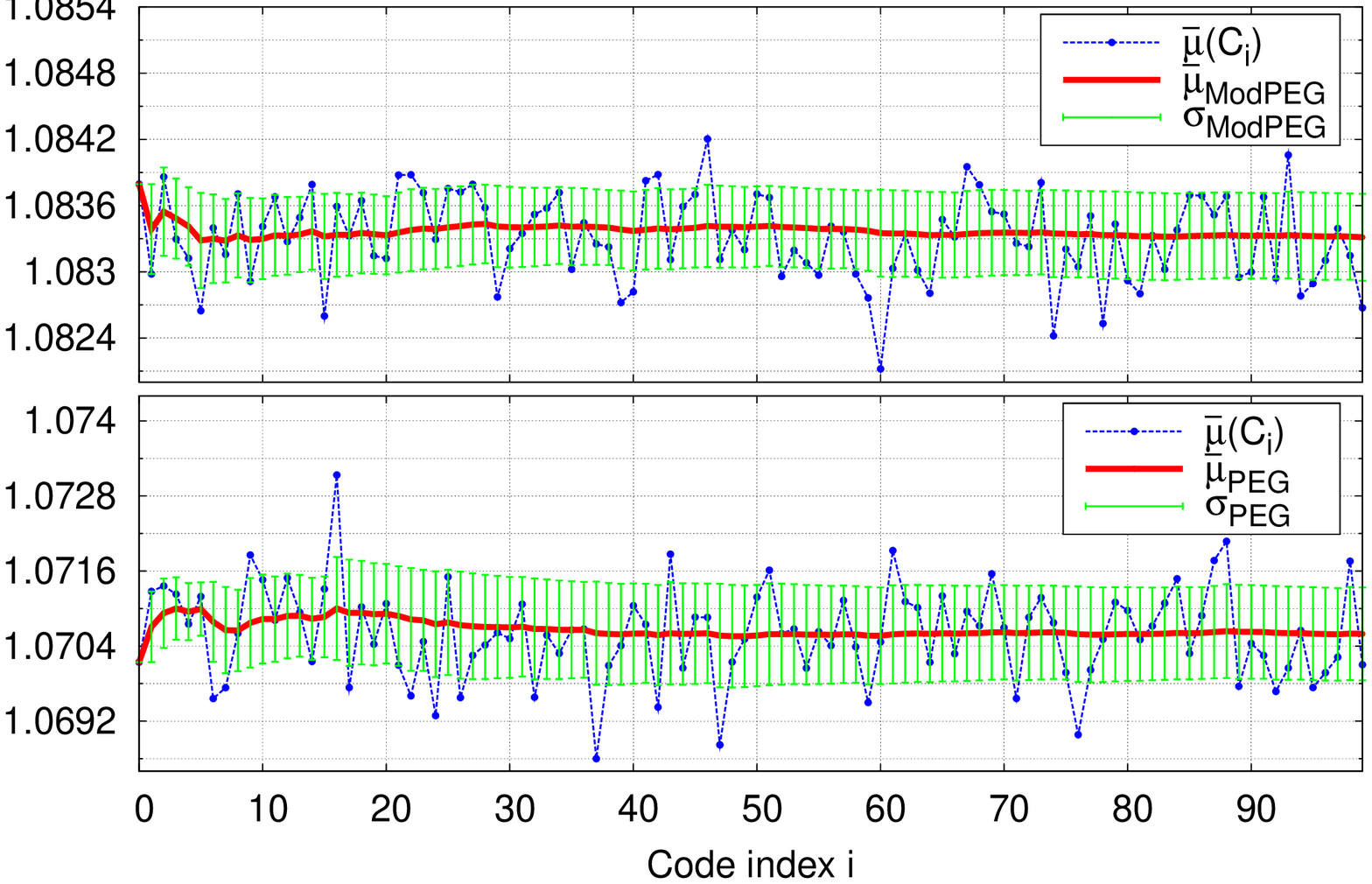}}
  \caption{Average inefficiency ratios of ensemble Tanner graphs constructed by using optimized Scheduled PEG algorithm in comparison with original PEG and modified PEG algorithms for $K = 5000$ information bits}
  \label{fig:peg_avg}
\end{figure*}

\begin{figure}[!t]
\centering
\includegraphics[width=\columnwidth]{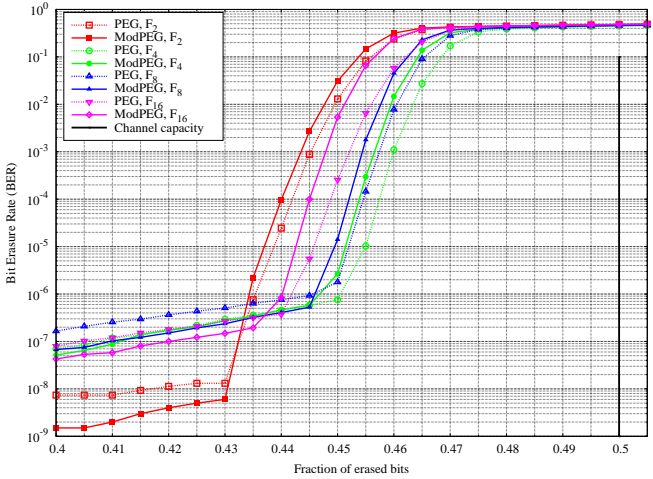}%
\caption{BER performance of Tanner graphs constructed by using PEG and ModPEG algorithms, for irregular
LDPC codes shown in Table \ref{tab:opt_irr_distr}, with with $K = 5000$ bits.}%
\label{fig:peg_ber_expl}%
\end{figure}

\begin{expl}
\label{expl:fin_perf_vs_asymp_perf} We consider irregular LDPC codes of rate $1/2$, defined over the $\f_{2}, \f_{4},
\f_{8}, \f_{16}$. The node-perspective symbol-node degree distributions and the asymptotic performance are shown in
Table \ref{tab:opt_irr_distr} (the constraint-node degree distributions are considered to be almost uniform). These
codes have been optimized by using density evolution methods. The binary code can be found in \cite{Shokrollahi2000b},
while non-binary codes have been optimized within this work. For each degree distribution, $100$ Tanner graphs have
been constructed, by both PEG and ModPEG algorithms, for codes with binary dimension $K = 5000$. The corresponding
average inefficiency ratios are also shown in Table \ref{tab:opt_irr_distr}. The details of these inefficiency ratios
are shown in Figure \ref{fig:peg_avg}, where it can be seen that using $\approx 20$ graphs is sufficient for obtaining
good estimates of $\inefpeg$ and $\inefmpeg$ values.

We also observe that, in all the four cases, the average inefficiency ratios $\inefpeg$ and $\inefmpeg$ are much larger
than $\ainef$. Also, the average inefficiency ratios corresponding to ModPEG  are larger than those corresponding to
the original PEG. The reason is that the ModPEG algorithm improves the error floor region but also worsens the
waterfall region, as it can be seen in Figure \ref{fig:peg_ber_expl}: there are eight BER curves, corresponding to
eight LDPC codes drawn from the corresponding PEG and ModPEG ensembles of graphs.
\end{expl}

\subsection{Scheduled Progressive Edge Growth algorithm}

Both PEG and ModPEG algorithms construct Tanner graph with large girth and low error floor. For optimized irregular
LDPC codes, however, the average inefficiency of finite length graphs is far away from the predicted (asymptotical)
threshold (even for codes with $K = 5000$ and $N = 10000$ bits, cf. Example \ref{expl:fin_perf_vs_asymp_perf}).

The Scheduled Progressive Edge Growth (SPEG) algorithm, proposed in this section, aims to improve the average
inefficiency of irregular LDPC codes.

We fix an integer $T \geq 1$. We consider a collection of disjoint symbol-nodes subsets $\vsdt\subseteq S$, indexed by
$t\in\{1,\dots,T\}$ and $d\in\{1,\dots\dm\}$, such that:
\begin{itemize}
  \item $\vsdt \subseteq S_d$ ($\vsdt$ contains only symbol-nodes of degree $d$)
  \item $\vs = \displaystyle \bigcup_{d=1}^{\dm} \bigcup_{t=1}^{T} \vsdt$.
\end{itemize}
We denote by $\ndt$ the number of symbol-nodes in $\vsdt$. It follows that $S_d = \cup_{t=1}^{T}\vsdt$, $n_d =
\displaystyle \sum_{i=0}^{t-1}\ndt$, and $n = \displaystyle \sum_{d=1}^{\dm} \sum_{i=0}^{t-1}\ndt$

The scheduled PEG algorithm works as follows: at time $t$, it connects progressively symbol-nodes within the subsets
$\vsdt$, for $d=1,\dots,\dm$. For each subset $\vsdt$, edges are established in a {\em degree-by-degree} manner.

\begin{algorithm}
\caption{Scheduled Progressive Edge Growth algorithm}
\label{algo:SPEG}
\begin{algorithmic}
\FOR{$t$ = 1 \TO $T$}
    \FOR{$d$ = 1 \TO $\dm$}
        \FOR{$k$ = 1 \TO $d$}
            \FOR{$s_j \in \vsdt$}
               \STATE Determine $\underline{C}_{s_j}$, given the current $E$;
               \STATE $c_i \longleftarrow \left\{\underline{C}_{s_j} \mid \mbox{min deg}\right\}$;
               \STATE Add edge $(s_j, c_i)$ to $E$;
            \ENDFOR
        \ENDFOR
    \ENDFOR
\ENDFOR
\end{algorithmic}
\end{algorithm}

We denote by $\speg$ the ensemble of all Tanner graphs constructed by using the SPEG algorithm. The average
inefficiency ratio over all graphs in $\speg$ is defined as:
$$\inefspeg\!\inputdataSPEG = \text{E}\!\left[ \inefm(\code) \mid \code \in \speg \right]$$
When parameters $\inputdataSPEG$ are implied, it will be simply denoted by $\inefspeg$. The corresponding standard
deviation is denoted by $\stdspeg$.

In the simple case of $T = 1$, we  have $S_d^{(1)} = S_d$, and the SPEG algorithm is equivalent to the ModPEG
algorithm. On the other hand, if $T = n$, each subset $\vsdt$ contains one single symbol-node, and the SPEG algorithm
is equivalent to the original PEG algorithm. For a general $T$ value, the SPEG algorithm is in between these two
extreme cases. It allows to explore the ensemble of LDPC codes with {\em fixed code-length and degree distributions}
and, as explained shortly, is aimed at finding codes with very small average inefficiency. As observed in Section
\ref{subsec:perf_metrics}, a lower average inefficiency corresponds to better performance of the code in the waterfall
region. However, generally there is a tradeoff between waterfall and error floor regions. Hence, to prevent excessive
degradation in the error floor region, the edges within each $\vsdt$ subset are established in a {\em degree-by-degree}
manner.


The use of the SPEG algorithm can potentially improve  the average inefficiency of the constructed LDPC code, as shown
in the following example.
\begin{expl}\label{expl:speg_potential}
We consider non-binary LDPC codes with rate $1/2$, defined over $\f_{16}$, whose node-perspective degree distribution
polynomials are given in Table \ref{tab:opt_irr_distr}. The asymptotic threshold is equal to $\pth = 0.4945$,
corresponding to an asymptotic inefficiency $\ainef = 1.011$.

We want to construct a code with binary dimension $K = 5000$. Hence, the Tanner graph contains $n = 2500$ symbol-nodes
and $m = 1250$ constraint-nodes. We use the SPEG algorithm with $T = 3$, such that $n_d^{(t_1)} = n_d^{(t_2)}$, for any
$1\leq t_1,t_2 \leq T$. This means that each $S_d$ is partitioned into three subsets $\vsdt, t=1,2,3$, of same
cardinality.

As in Example \ref{expl:fin_perf_vs_asymp_perf}, we estimated the average inefficiency ratio of the ensemble $\speg$ by
simulating $100$ SPEG-codes. We obtained $\inefspeg = 1.0536$, which has to be compared with $\inefpeg = 1.0706$ and
$\inefmpeg = 1.0834$ from Table \ref{tab:opt_irr_distr}.
\end{expl}

This example shows that an appropriate choice of the subsets $\left\{\vsdt\right\}$ may improve the average
inefficiency of the constructed code. In the following section we propose a method that allows to optimize this choice,
by minimizing the average inefficiency of the corresponding ensemble of SPEG-codes.

\section{Optimized Scheduled-PEG construction}\label{sec:Opt_SPEG}
\subsection{Optimization algorithm}

The main idea behind the SPEG algorithm is that different choices of the {\em scheduling subsets}
$\left\{\vsdt\right\}$ might lead to codes with different performance. Our purpose is to find scheduling subsets that
minimize the average inefficiency of the corresponding ensemble of SPEG-codes. In order to properly formulate this
optimization problem, we have to take into consideration only the ``profile'' of the scheduling subsets, which consists
of the fractions of nodes within each subset. Precisely, let $\fsdt$ denote the fraction of symbol-nodes contained in
$\vsdt$; hence:
\begin{enumerate}
    \renewcommand{\theenumi}{(\arabic{enumi}}
    \item $\fsdi = \displaystyle\frac{\ndi}{n}$
    \item $\displaystyle \sum_{t=1}^{T}\fsdt = \Lambda_d$
    \item $\displaystyle \sum_{d=1}^{\dm}\sum_{t=1}^{T}\fsdt = \sum_{d=1}^{\dm}\Lambda_d=1$
\end{enumerate}
A family of parameters $\left\{\fsdt\right\}_{\mbox{\!\!\scriptsize$\begin{array}{l} 1\leq t \leq T \\
1 \leq d \leq \dm\end{array}$\!\!\!\!\!\!\!\!\!}}$ satisfying the condition (2) before is called {\em scheduling
distribution}.

When parameters $(n, m, \left\{\fsdt\right\})$ are fixed, we consider that the SPEG algorithm starts by randomly
choosing a family of scheduling subsets $\left\{\vsdt\right\}$, according to the given scheduling distribution, and
then it constructs a (random) Tanner graph as explained in the above section. We denote by $\spegopt$ the corresponding
ensemble of SPEG Tanner graphs, and we define its average inefficiency by:
$$\inefspeg\!\inputdataSPEGopt = \text{E}\!\left[ \inefm(\code) \mid \code \in \spegopt \right]$$
When parameters $n$ and $m$ are fixed, it will be simply denoted by $\inefspeg\left(\left\{\fsdt\right\}\right)$. This
represents the objective function of our optimization problem. Although it cannot be computed analytically,
$\inefspeg\left(\left\{\fsdt\right\}\right)$ can be efficiently estimated by simulating a finite number of codes $\code
\in \spegopt$.

\begin{figure}[!t]
\centering
\includegraphics[width=\columnwidth]{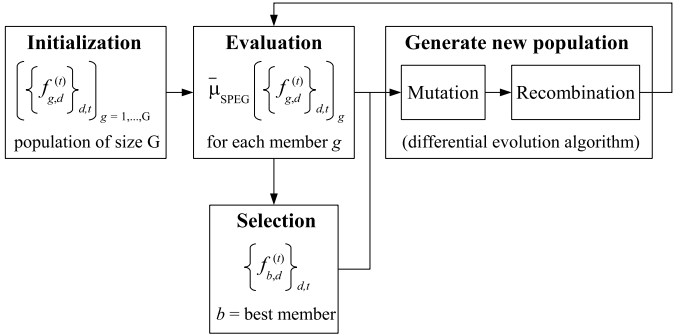}%
\caption{Flowchart of the differential evolution optimization algorithm}%
\label{fig:speg_flowchart}%
\end{figure}

Since our objective function cannot be expressed analy\-tically, we address the optimization problem by using generic
population-based metaheuristic optimization algorithms. More precisely, the {\em differential evolution} algorithm
\cite{DiffEvol} is used, which optimizes $\inefspeg\left(\left\{\fsdt\right\}\right)$ by iteratively trying to improve
the best current solution $\left\{f_{b,d}^{(t)}\right\}$. The flowchart of the optimization algorithm is illustrated in
Figure \ref{fig:speg_flowchart}.

The algorithm maintains a population of candidate solutions, which is randomly initialized. At each iteration, the
current population is evaluated, such as to find the best current solution. Then, a population of new candidate
solutions is generated by combining existing candidates (mutation and recombination), and then keeping whichever
candidate solution has the best score. In this way the objective function is treated as a black box that provides a
measure of quality of the candidate solutions.

\subsection{Simulation results}

We consider four ensembles of irregular LDPC codes of rate $1/2$, defined over the $\f_{2}, \f_{4}, \f_{8}, \f_{16}$,
whose node-perspective symbol-degree distributions are shown in Table \ref{tab:opt_irr_distr} (the constraint-node
degree distributions are considered to be almost uniform).

\begin{figure}[!t]
\centering
\includegraphics[width=\columnwidth]{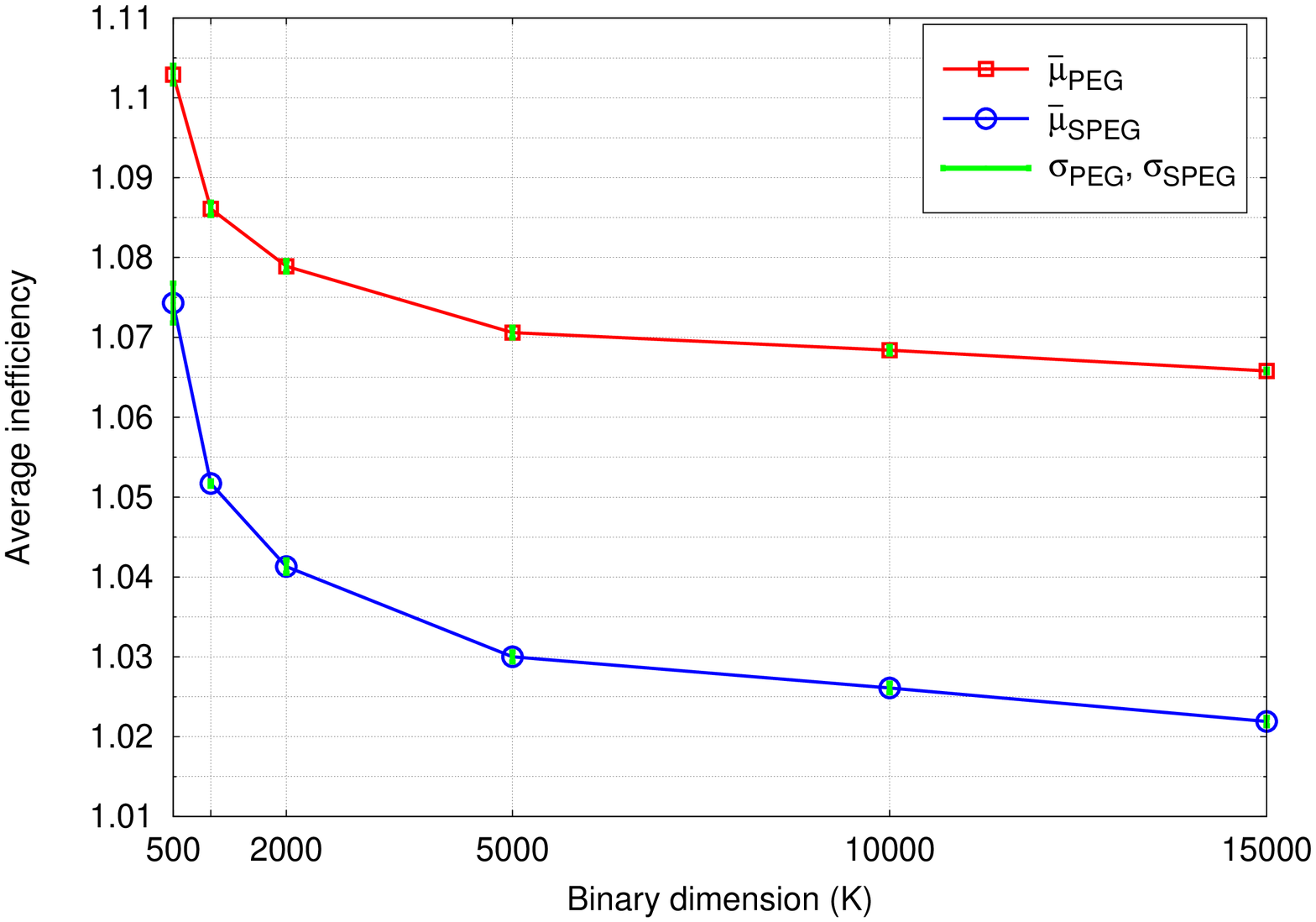}%
\caption{Average inefficiency ratios of the ensembles $\peg$ and $\spegopt$}%
\label{fig:ineff_ratio}%
\end{figure}

\begin{figure}[!t]
\centering
\includegraphics[width=\columnwidth]{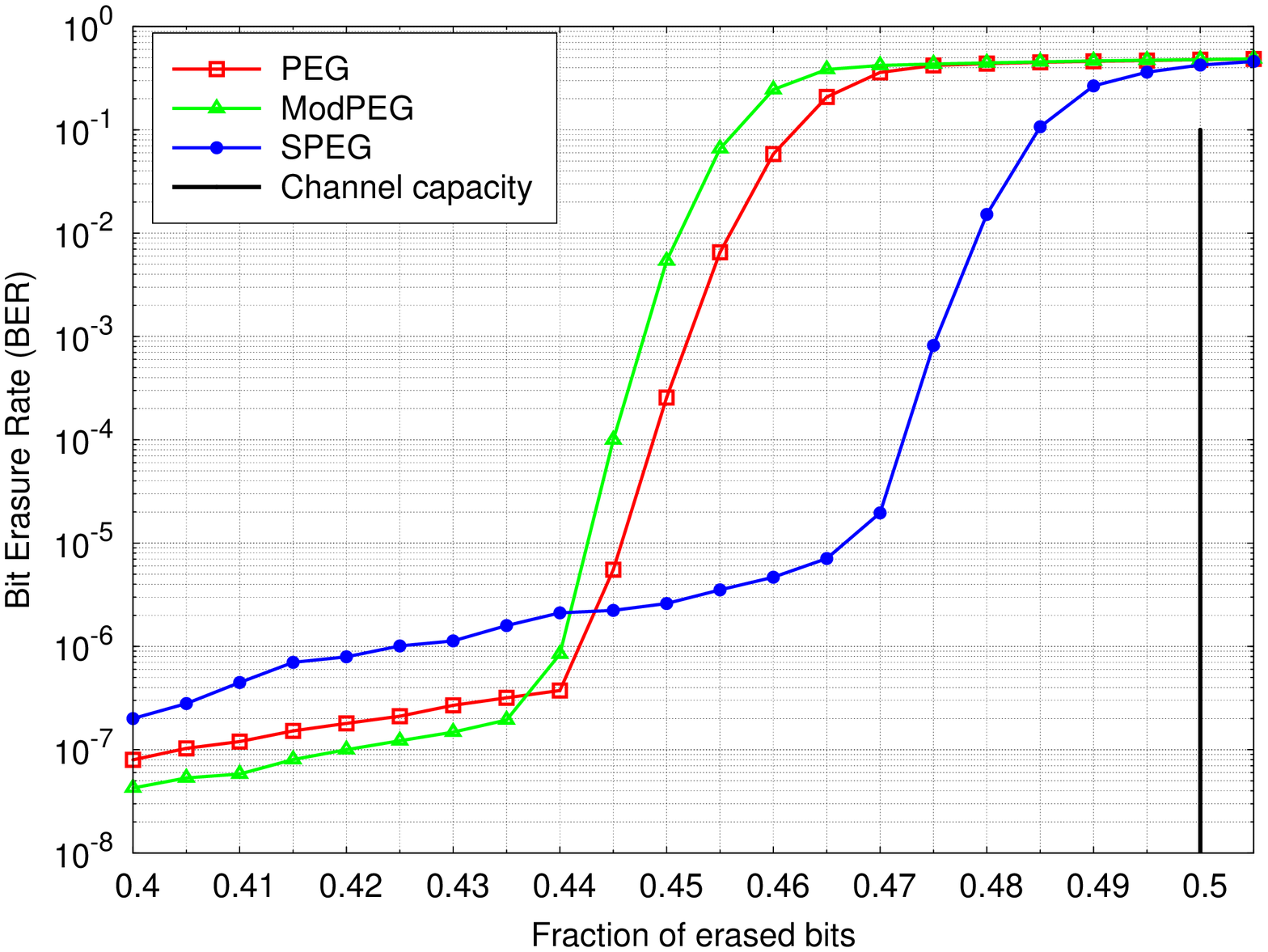}%
\caption{Bit Erasure Rates of three (among the best) codes from the ensembles $\peg$, $\mpeg$, and $\spegopt$}%
\label{fig:speg_ber_fer}%
\end{figure}

\begin{table*}[!t]
\centering \caption{Optimized scheduling distributions and corresponding inefficiency ratios, for irregular LDPC codes
of rate $1/2$, defined over $\f_2$, $\f_4$, $\f_8$ and $\f_{16}$, with binary dimension $K=5000$} \label{tab:opt_simul}
\begin{tabular}{|c|l|r|r||r|r||r|}
\hline
Alphabet & Optimized Scheduling & $\stdspeg$ & $\inefspeg$  & $\inefpeg$ & $\inefmpeg$ & $\inefrand$\\
\hline \hline
  $\f_2$ & $\begin{array}{@{\!\!}|c|c|c|c|c|@{\!\!}}
t &  f_{2}^{(t)} &  f_{3}^{(t)} &  f_{7}^{(t)} &  f_{30}^{(t)} \\
\hline
1 & 0.2939 & 0.0690 & 0      & 0.0071 \\
\cline{1-1}
2 & 0.2523 & 0.1797 & 0.0787 & 0.0223  \\
\cline{1-1}
3 & 0.0028 & 0.0018 & 0.0820 & 0.0104 \\
\end{array}$
& 7.7e-04 & 1.0326  & 1.0829 & 1.0876 & 1.3367 \\
\hline \hline
  $\f_4$ & $\displaystyle\begin{array}{@{\!\!}|c|c|c|c|@{\!\!}}
t &  f_{2}^{(t)} &  f_{4}^{(t)} &  f_{12}^{(t)} \\
\hline
1 & 0.1876  & 0.0099    & 0.0040 \\
\cline{1-1}
2 & 0.2088  & 0       & 0.0021 \\
\cline{1-1}
3 & 0.3175  & 0.2073    & 0.0626 \\
\end{array}$
& 3.301e-04 & 1.0384  & 1.0604  & 1.0685 & 1.3129 \\
\hline \hline
  $\f_8$ & $\begin{array}{@{\!\!}|c|c|c|c|c|@{\!\!}}
t &  f_{2}^{(t)} &  f_{3}^{(t)} &  f_{5}^{(t)} &  f_{12}^{(t)} \\
\hline
1 & 0.1120 & 0.0028 &   0.0020  & 0.0065 \\
\cline{1-1}
2 & 0.1844 & 0.0084 & 0.0036    & 0.0026 \\
\cline{1-1}
3 & 0.4893 & 0.0417 & 0.1097    & 0.0371  \\
\end{array}$
& 3.049e-04 & 1.0327  & 1.0655 &    1.0726 & 1.2256 \\
\hline \hline
  $\f_{16}$ & $\begin{array}{@{\!\!}|c|c|c|c|c|@{\!\!}}
t &  f_{2}^{(t)} &  f_{5}^{(t)} &  f_{8}^{(t)} &  f_{18}^{(t)} \\
\hline
1 & 0.4118  & 0.0473    & 0             & 0.0004 \\
\cline{1-1}
2 & 0.1662  & 0.0004    & 0.0002    & 0.0071 \\
\cline{1-1}
3 & 0.2680  & 0.0579    & 0.0023    & 0.0157  \\
\end{array}$
& 6.407e-04 & 1.0302 & 1.0706 & 1.0834 & 1.1905 \\
\hline
\end{tabular}
\end{table*}

We fixed the parameter $T=3$ and, for each of the above ensembles, we used the differential evolution algorithm to find
a scheduling distribution that minimizes the average inefficiency of the corresponding SPEG codes. We considered codes
with binary dimension $K = 5000$; Hence, the binary code-length is $N=10000$ and the non-binary code-length (number of
symbol-nodes of the graph) is $n=N/p$, where $p\in\{1,2,3,4\}$. The optimized scheduling distributions
$\left\{\fsdt\right\}$ and the corresponding inefficiencies $\inefspeg\left(\left\{\fsdt\right\}\right)$ are shown in
Table \ref{tab:opt_simul}. For comparison purposes, we have also displayed in Table \ref{tab:opt_simul} the average
inefficiencies of the corresponding PEG, ModPEG, and random\footnote{Random codes with given $n$, $m$, and node-degree
distribution polynomials.} ensembles. We can see that the SPEG algorithm significantly improves the average
inefficiency ratios in comparison with those of the PEG and ModPEG algorithms.

Moreover, we constructed several finite length codes over $\f_{16}$, using the same scheduling distribution that has
been optimized for $K=5000$. The average inefficiency of these codes is plotted in Figure \ref{fig:ineff_ratio}. It can
be observed that they significantly outperform codes constructed by the original PEG algorithm.

For $K=5000$, Figure \ref{fig:speg_ber_fer} displays the bit error rates of one PEG code, one ModPEG code, and one SPEG
code with optimized scheduling distribution (these codes are chosen among the best codes of the corresponding
ensembles). We can see that the SPEG algorithm significantly improves the waterfall region, at the expense of a
slightly higher error floor.

 \section{Conclusion}\label{sec:Conclusion}

The proposed Scheduled-PEG algorithm allows the enhancement of the classical PEG algorithm, by the introduction of a
scheduling distribution that specifies the order in which edges are established in the graph. The scheduling
distribution provides a way for exploring the ensemble of LDPC codes with fixed code-length and degree distributions,
and is aimed at finding codes with very small average inefficiency.

We showed that the SPEG algorithm can be successfully combined with genetic optimization algorithms, which
significantly improves the average inefficiency of the constructed LDPC codes over the classical-PEG construction. In
terms of error rate curves, this translates into a significant improvement of the waterfall region.

Finally, we remark that the optimization of the scheduling distribution makes use of the specific channel model
(through the use of the decoding inefficiency). Hence, LDPC codes constructed by using the SPEG algorithm together with
an optimized scheduling distribution are channel dependent. However, the proposed algorithm could be generalized for
more general channel models (e.g. by optimizing with respect to a target FER, or to the area under the FER curve.).

\bibliographystyle{./Bibliography/IEEEbib}
\footnotesize
\bibliography{./Bibliography/ECR,./Bibliography/savin}

\end{document}